\documentclass[fleqn,10pt]{wlscirep}
\usepackage[utf8]{inputenc}
\usepackage[T1]{fontenc}
\usepackage{amsmath}
\usepackage{xcolor}
\usepackage[font={sf,small}]{caption}
\usepackage{xfrac}
\usepackage[inkscapelatex=false]{svg}

\title{High-Precision Fluidic Kirigami Metasurface for Ultrasonic Holographic Lensing and Haptic Interfacing}

\author[1,+,*]{Ardalan Kahak}
\author[1,+]{Moustafa Sayed Ahmed}
\author[1,+]{Nahid Kalantaryardebily}
\author[1,]{Hrishikesh Kulkarni}
\author[1,2,]{Netta Gurari}
\author[1,]{Shima Shahab}
\author[1,*]{Suyi Li}

\affil[1]{Virginia Tech, Mechanical Engineering, Blacksburg, 24061, USA}
\affil[2]{Virginia Tech, Biomedical Engineering and Mechanics, Blacksburg, 24061, USA}

\affil[*]{corresponding authors: ardalankahak@vt.edu, suyili@vt.edu}

\affil[+]{these authors contributed equally to this work}

\begin{abstract}
Morphing surfaces provide a versatile tool to advance the functionalities of high-performance aircraft, soft robots, biomedical devices, and human-machine interfaces. However, achieving \textit{precise} shape transformation and mechanical property control remains challenging due to nonlinearity, design constraints, and the difficulty of coordinating multiple constituent materials across a continuous surface. To this end, this study unveils that Kirigami art can inspire novel solutions. More specifically, the geometric principles of Kirigami can be exploited to design and fabricate fluidic metasurfaces capable of highly accurate shape morphing and output force control, all via a single pressure input. 
This study presents a systematic approach to designing Kirigami for tuning the local deformation and force output through extensive nonlinear modeling and experiment validation on two archetypal patterns: concentric square and circular cuts. 
The potentials of this approach are demonstrated via two multiphysics case studies: (1) an acoustic holography lens for ultrasonic wave steering, achieving highly accurate deformation control under a single global pressure input, and (2) a haptic device with a small volume, constant contact area, and high-resolution output force. 
The fluidic Kirigami concept allows for simple yet effective adaptation to different shape and stiffness requirements, paving the way for a new family of scalable and programmable morphing surfaces.


\end{abstract}

\begin{document}

\flushbottom
\maketitle

\thispagestyle{empty}

\section*{Introduction}

\begin{figure}
    \centering
    \includegraphics[scale=1.0]{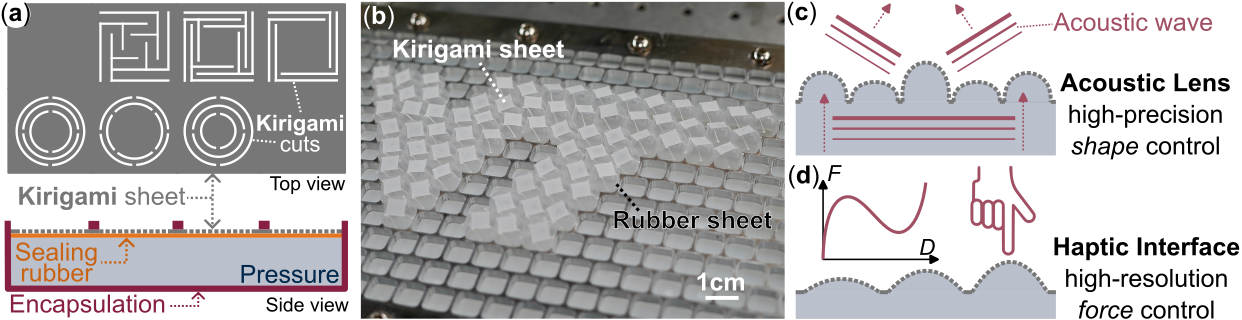}
    \caption{
        \textbf{The overall concept of fluidic Kirigami metasurface}. (a) By wrapping Kirigami cut sheets outside a fluid-fluid container, we can create a morphing surface. (b) Our proof-of-concept tests showed that fluidic pressure can deform the Kirigami into a 3D topology, and the design of the Kirigami dictates such deformation. Therefore, we can leverage the correlation between design and mechanics behind the Kirigami cuts to achieve high-precision shape control (e.g., for acoustic focusing shown in (c)) or high-resolution force control (e.g., for haptic interface shown in (d)).}
    \label{fig:vision}
\end{figure}

Soft and morphing surfaces have shown exciting capabilities to create and augment the interfaces between widely different physical domains \cite{wang2024performance, yang2023morphing} --- between fluid and solids \cite{thill2008morphing, barbarino2011review, ajanic2020bioinspired}, human and machines \cite{johnson2023multifunctional, qamar2018hci}, or bodies and their environments \cite{bai2022dynamically, pikul2017stretchable, murashima2021active}. By transforming its external shape or internal architecture, these morphing interfaces can be used to engineer critical interface behaviors like wave propagation \cite{venkatesh2022origami, assouar2018acoustic, zhang2021reconfigurable, tian2019programmable, cao2021tunable}, haptic feedback \cite{bai2021elastomeric, bau2009bubblewrap}, boundary flow development \cite{kang2020lock, thompson2022surface, shinde2021supersonic}, and object manipulation \cite{foresti2013morphing, liu2021robotic, wang2023passively, park2019bio, an2024mechanically}.
As a result, morphing surfaces are among the key drivers behind recent advances in soft robotics, adaptive acoustics, aerospace, and biomedical industries.
However, controlling the topology and mechanics of soft morphing surface \emph{with high precision} remains an open challenge.  In this regard, we present a solution by harnessing the out-of-plane deformation mechanics of Kirigami (``cut-paper'' in Japanese). The key idea here is to wrap an automatically designed, simple-to-fabricate Kirigami sheet outside a soft and fluid-filled container, creating a ``fluidic Kirigami metasurface'' that can deform into a pre-programmed 3D topology under a global pressure input (Fig.~\ref{fig:vision}a,b).
Kirigami cutting is a reliable and scalable fabrication approach. It can be implemented on a wide selection of thin-sheet precursors of different sizes, including graphene \cite{blees2015graphene, shi2024highly, moura2024ballistic}, electronics \cite{yang2024kirigami, li2019kirigami, zhuo2023kirigami, wei2024revolutionizing}, adhesives \cite{zhao2018kirigami, hwang2023metamaterial, hwang2018kirigami}, and metals \cite{morikawa2018ultrastretchable, liu2018nano}. Cutting can be achieved with very high precision and almost limitless design possibilities \cite{chen2020kirigami, zhang2022kirigami, sun2021geometric}. More importantly, Kirigami's mechanical behavior --- such as its external deformation under load, auxetics, and stiffness --- is directly dictated by the underpinning cut pattern design \cite{tao2023engineering,isobe2016initial, chaudhary2021geometric}. 
Therefore, the nonlinear mechanics of the Kirigami sheet offer a reliable mechanism to precisely control the shape and elastic properties of the morphing surface.

This study demonstrates the high-precision outputs of fluidic Kirigami metasurface via two inter-disciplinary functions. 
\emph{The first function is acoustic holographic lensing, which presents the most demanding shape control requirements} (Fig.~\ref{fig:vision}c). By carefully designing the Kirigami cut pattern and its corresponding 3D deformation under pressure, the fluidic Kirigami metasurface can transform into an acoustic holographic lens, focusing ultrasonic waves in a targeted direction. If a new acoustic focusing pattern is needed, one only needs to peel off the Kirigami sheet and attach a new one with a different cut pattern. Accurate manipulation of acoustic waves, especially ultrasonic ones for medical applications, requires precise deformation of the metasurface within the wavelength (typically in the millimeter range)~\cite{wu2024wave,imani2024advanced}. For this reason, acoustic lensing is an ideal case for validating the high-precision shape control on the morphing surface using the Kirigami principle.

\emph{The second function is the haptic interface, which presents a demanding high-resolution force control requirement} (Fig.~\ref{fig:vision}d). Precise, adaptable, and customizable haptic feedback devices require force-deformation relationships tailored to desired human muscular and sensory preferences while situated in potentially small areas (e.g.,~below one's fingers or toes). This precise force feedback for targeted displacements is crucial in applications such as virtual reality, robotic surgery, and human touch-related neurophysiological investigations~\cite{paul_scoping_2024,bermejo_survey_2022,patel_haptic_2022,el_rassi_review_2020}. For each of these applications, high-resolution force output may be needed with minimal displacement to convey realistic physical interactions to a user. The inclusion of fluidic Kirigami enables such interactions, given its ability to deform out-of-plane in response to fluidic changes, such as air, with a mathematically modeled stiffness. An additional benefit of haptic devices created using fluidic Kirigami is that the design can be relatively small in size, with the physical interaction arising due to small inflation, followed by deflation of an object (e.g.,~rubber membrane of a small balloon/bubble).

Achieving these two widely different functions with high precision requires mechanics modeling that can accurately correlate the Kirigami cut pattern design with its deformation and force output. Critically, this analytical model is the prerequisite for automatic and efficient design. Therefore, the following part of this paper will start by detailing the new nonlinear mechanics model of the Kirigami cell, then explain the inverse design process and demonstrate its use in acoustic lensing and haptic interfacing. In particular, we will use a square-cut pattern for the acoustic lensing because it allows significant out-of-plane deformation, and circular-cut for haptic feedback because it provides more precise force control. 

\section*{Square-cut Kirigami Surface for High-Precision Shape Control}
The square-cut Kirigami cell consists of concentric ``rings'' of straight and interconnected thin ligaments of length $l$ and width $w$ (Fig.~\ref{fig:Scut}a). Therefore, a square-cut Kirigami cell can be fully defined by the following geometric parameters: the unit cell size $d$, the number of rings $N$, the ligament width $w$, and the gap between the cuts $g$. Notice the length $l$ of each ligament can be directly calculated based on these parameters. 

\subsection*{Analytical tool for designing the square-cut Kirigami cells' deformation}
\begin{figure}[b!]
    \centering
    \includegraphics[scale=1.0]{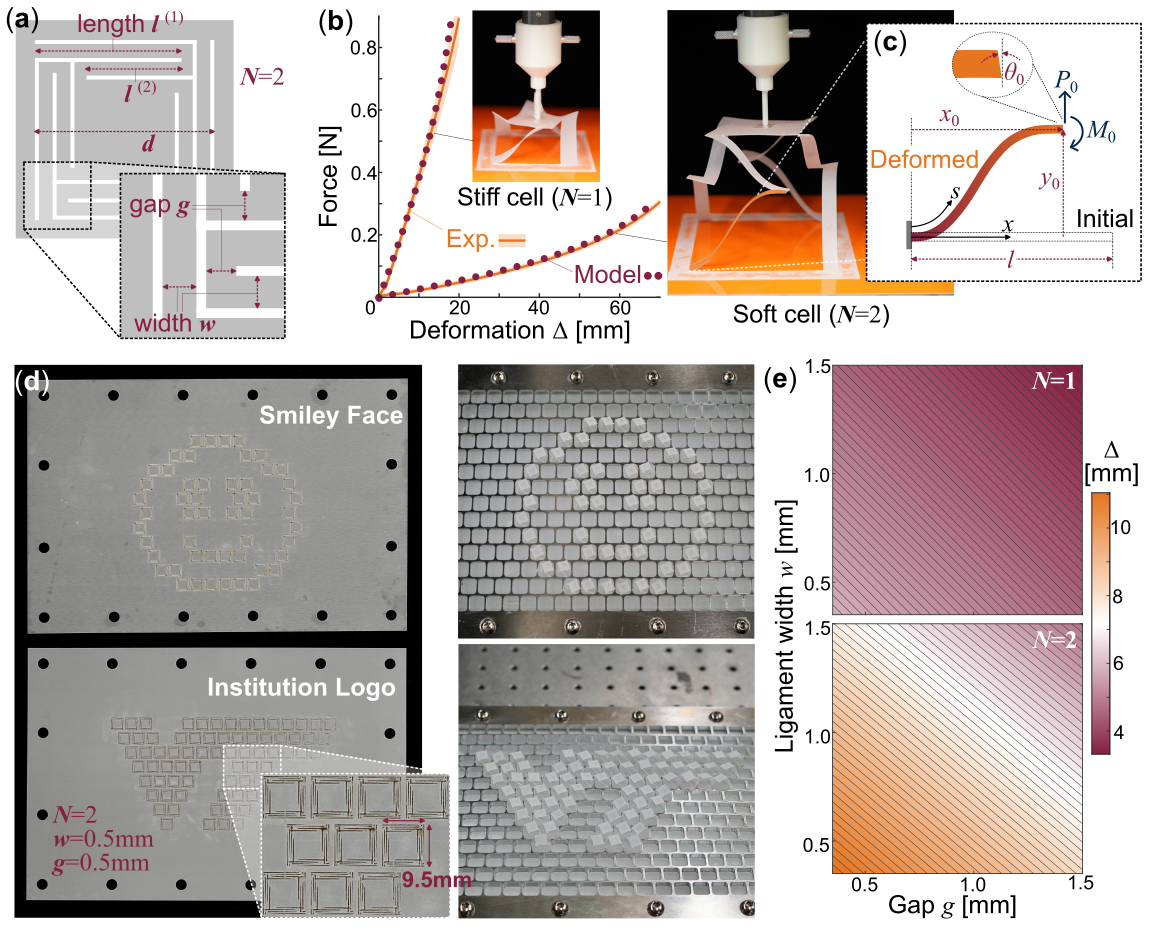}
    \caption{\textbf{The anatomy and design space of a square-cut fluidic Kirigami metasurface}. (a) An example design with two concentric rings of straight ligaments ($N=2$). Here, we use superscript $^{(N)}$ to denote the different rings. (b) The out-of-plane deformation response of two square-cut Kirigami cells. The solid lines are averaged test results from 10 consecutive loading cycles, and the shaded bands are the corresponding standard deviation. The dotted lines are the analytical model prediction. (c)~The deformation kinematics of an individual ligament modeled as a nonlinear Euler-Bernoulli beam. (d) A demonstration of pre-programmed Kirigami sheets ``displaying'' two different patterns. One can simply swap a new Kirigami sheet if a new pattern is desired. (e) A parametric study showing the design space of 9.5mm$\times$9.5mm square-cut Kirigami unit cells subject to 60kPa fluidic pressure, based on the analytical solution defined in Eq. (\ref{eq:Delta}).}
    \label{fig:Scut}
\end{figure}

In this study, we focus on Kirigami cells with only one or two rings of ligaments ($N=1,2$) and model each ligament as a straight cantilever beam with compound loading at the free end (Fig.~\ref{fig:Scut}b,c). The \textit{Euler-Bernoulli beam theory} describes their finite and geometrically non-linear deflection by:
\begin{equation}   \label{Euler}
    M(x) = EI \frac{d \theta}{d s} = EI \frac{\frac{d^2 y}{d x^2}}{\left[1+\left(\frac{d y}{d x}\right)^2\right]^{3/2}},
\end{equation}
where $M(x)$ is the reaction bending moment along the beam span, $\theta$ is the beam cross-sections' rotation angle, $s$ is the position along the deformed beam arc, $x$ is the position along the initial, non-deflected beam, and $y$ is the out-of-plane beam deflection. 
Assuming the nonlinear cantilever beam is subject to an external, transverse force $P_0$ and bending moment $M_0$ at its tip, Eq. (\ref{Euler}) can be re-formulated into\cite{kimball2002modeling}:
\begin{equation} \label{Non-linear}
\begin{aligned}
\alpha  &= \pm \frac{1}{\sqrt{2}} \int_0^{\theta_0} \frac{d \theta}{\sqrt{\lambda-\sin \theta}},\\
\beta_x &= \frac{x_0}{l} = \pm \frac{1}{\alpha \sqrt{2}} \int_0^{\theta_0} \frac{\cos \theta \, d \theta}{\sqrt{\lambda-\sin \theta}} ,\\
 \beta_y  &=\frac{y_0}{l}  = \pm \frac{1}{\alpha \sqrt{2}} \int_0^{\theta_0} \frac{\sin \theta \, d \theta}{\sqrt{\lambda-\sin \theta}},
\end{aligned}
\end{equation}
where $\alpha$ is a non-dimensional force variable. $x_0$ and $y_0$ are the horizontal and vertical beam displacements at its tip, respectively, so $\beta_x$ and $\beta_y$ are normalized tip deflections. $\theta_0$ is the beam's rotation at the tip (Fig.~\ref{fig:Scut}c).  The $\pm$ sign is positive (or negative) if the beam slope monotonically increases (or decreases). $\lambda$ is a non-dimensional angle variable in that:
\begin{equation} \label{lambda}
\begin{aligned}
    \lambda &= \sin\theta_0 + \kappa,\\
    \kappa &= \frac{1}{2} \frac{M_0^2}{P_0 EI}.\\
\end{aligned}
\end{equation}

The generic solution of Eq. (\ref{Non-linear}) when  \(0 \leq |\lambda| \leq 1\) is provided in Section 1 of the supplement information. By closely examining the deformation pattern of the square-cut Kirigami cell (Fig.~\ref{fig:Scut}b), we can assume that $P_0$ is positive, $M_0$ is negative, and $\theta_0 \approx 0$ (therefore, $\lambda \approx \kappa$ in Eq. (\ref{lambda})). Notice that if the external force is sufficiently high, $\theta_0$ will deviate away from 0. However, this $\theta_0 \approx 0$ assumption can provide us with accurate predictions without incurring unnecessary complex formulations.
Based on these assumptions, we can apply elliptic integral solutions to solve Eq. (\ref{Non-linear}) so that:
\begin{equation} \label{elliptic_equ}
\begin{aligned}
\alpha &= f^* ,\\
\beta_x &= \frac{x_0}{l} = \frac{2\sqrt{2\kappa}}{\alpha}  ,\\
\beta_y &= \frac{y_0}{l} = \frac{f^* - 2 e^* + 2\sqrt{2\kappa}}{\alpha},
\end{aligned}
\end{equation}
where \( f^* \), \( e^* \) are the complete elliptic integral of the first and second kind, respectively:
\begin{equation}
    \begin{aligned}
    f^*&=2F\left(\gamma, k\right) \equiv 2  \int_0^{\gamma} \frac{1}{\sqrt{1-\kappa^2 \sin ^2 \nu}} d \nu, \\
    e^*&=2E\left(\gamma, k\right) \equiv 2 \int_0^{\gamma} \sqrt{1-\kappa^2 \sin ^2 \nu} d \nu, 
    \end{aligned}
\end{equation}
and:
\begin{equation}
    \begin{aligned}
\gamma  &=\sin ^{-1} \sqrt{\frac{2 \kappa}{\kappa+1}}, \\
k &=\sqrt{\frac{\kappa+1}{2 }} .
    \end{aligned}
\end{equation}

In addition to the equations above, we can also apply an external force condition in that: 
\begin{equation}\label{force}
    \alpha = \sqrt{\frac{P_0l^2}{EI}},
\end{equation}
where $P_0 = p_0A/M$. Here, $p_0$ is the global fluidic pressure, $A$ is the Kirigami unit cell’s area, and $M$ is the number of cantilever beams in each layer ($M=4$ in the square-cut design). In this way, the only unknown left in Eq. (\ref{elliptic_equ}-\ref{force}) is the bending moment $M_0$, which a simple numerical search can solve. Then, one can use the second and third equations in  Eq. (\ref{elliptic_equ}) to solve the beams' normalized tip displacement $\beta_x$ and $\beta_y$. 

The solutions above are formulated for a single nonlinear cantilever beam (or Kirigami ligament), but they apply to any ligaments in the square-cut Kirigami cell. Denote $\beta_y^{(1)}$ and $\beta_y^{(2)}$ as the normalized beam tip displacement of rings (1) and (2), respectively (the outermost ring is labeled as (1)). The overall Kirigami cell displacement $\Delta$ is:  
\begin{equation}
\begin{aligned}
\Delta = \beta_{y}^{(1)}l^{(1)} + \beta_{y}^{(2)}l^{(2)}.
\end{aligned}\label{eq:Delta}
\end{equation}

This model's outcome is consistent with a previous work that only involves a single load $P_0$\cite{belendez2002largeand}. To assess the precision of this model, we cut two Kirigami unit cells with different designs and measured their out-of-plane stiffness on a universal tensile test machine ($0.19$mm thin color-coded plastic shim with $E\approx5.5$GPa and Instron 68TM-5 with 10N load cell). One unit cell sample is designed to be relatively stiff with $d=50$mm, $N=1$, $w=10$mm, and $g=5$mm, while the other one is soft with $d=80$mm, $N=2$, $w=8$mm, and $g=7.5$mm. They are both fixed to the tester table along its outer perimeter and subject to a point force at its middle via a 3D-printed customized adapter. Both Kirigami cells show consistent force-deformation curves through repeated loading cycles, and the test result agrees well with the analytical model prediction. Small discrepancies only occur when the external force is very high, when our $\theta_0 \approx 0$ assumption is no longer accurate. Regardless, this simple analytical model offers a powerful tool for designing the fluidic Kirigami metasurface for accurate shape morphing.

\subsection*{Unique advantages of using fluidic Kirigami for shape morphing} 
The first advantage is its modularity. A single global pressure input can activate all the Kirigami unit cells. Moreover, cutting is a simple and scalable fabrication procedure (e.g., using a laser cutter or mechanical plotter cutter) \cite{khatak2022laser, bartholomeusz2005xurography}. Therefore, one can quickly design and cut Kirigami sheets on demand according to the desired morphing output. To demonstrate this, we cut two Kirigami sheets --- each consisting of a uniform array of square-cut Kirigami unit cells --- to ``display'' two patterns like a pixel grid. One is a smiley face, and the other pattern is the logo of the authors' institution (Fig.~\ref{fig:Scut}d). This morphing surface consists of a thin rubber sheet for pressure proofing, the plastic Kirigami sheet, and 6mm thick aluminum support layer with square cutouts to prevent overall bulging from the pressurization (Section 2 of the supplement information and Movie S1 detail this test setup). With a global pressure input, each Kirigami cell will deform locally according to its underpinning design. Most importantly, if a new display pattern is desired, one can quickly swap Kirigami sheets with different designs.

The second advantage of the fluidic Kirigami metasurface concept is its large design space. By simply tailoring the cutting geometry, we can tailor the unit cell deformation considerably. For example, Fig.~\ref{fig:Scut}e shows the result of a parametric study based on the aforementioned analytical tool. In this study, we fix the unit cell size $d=9.5$mm and the number of concentric rings $N=1$ or $2$. The actuation pressure is set at 60kPa. By adjusting the cantilever beam width $w$ and gap size $g$, one can continuously prescribe the unit cell deformation from 0 to 14mm (a.k.a., \textbf{1.47} $\Delta/d$ aspect ratio). Therefore, it becomes possible to fine-tune the deformation of each unit cell accurately and independently within a large range.

\begin{figure}[h!]
    \centering
    \includegraphics[scale=1.08]{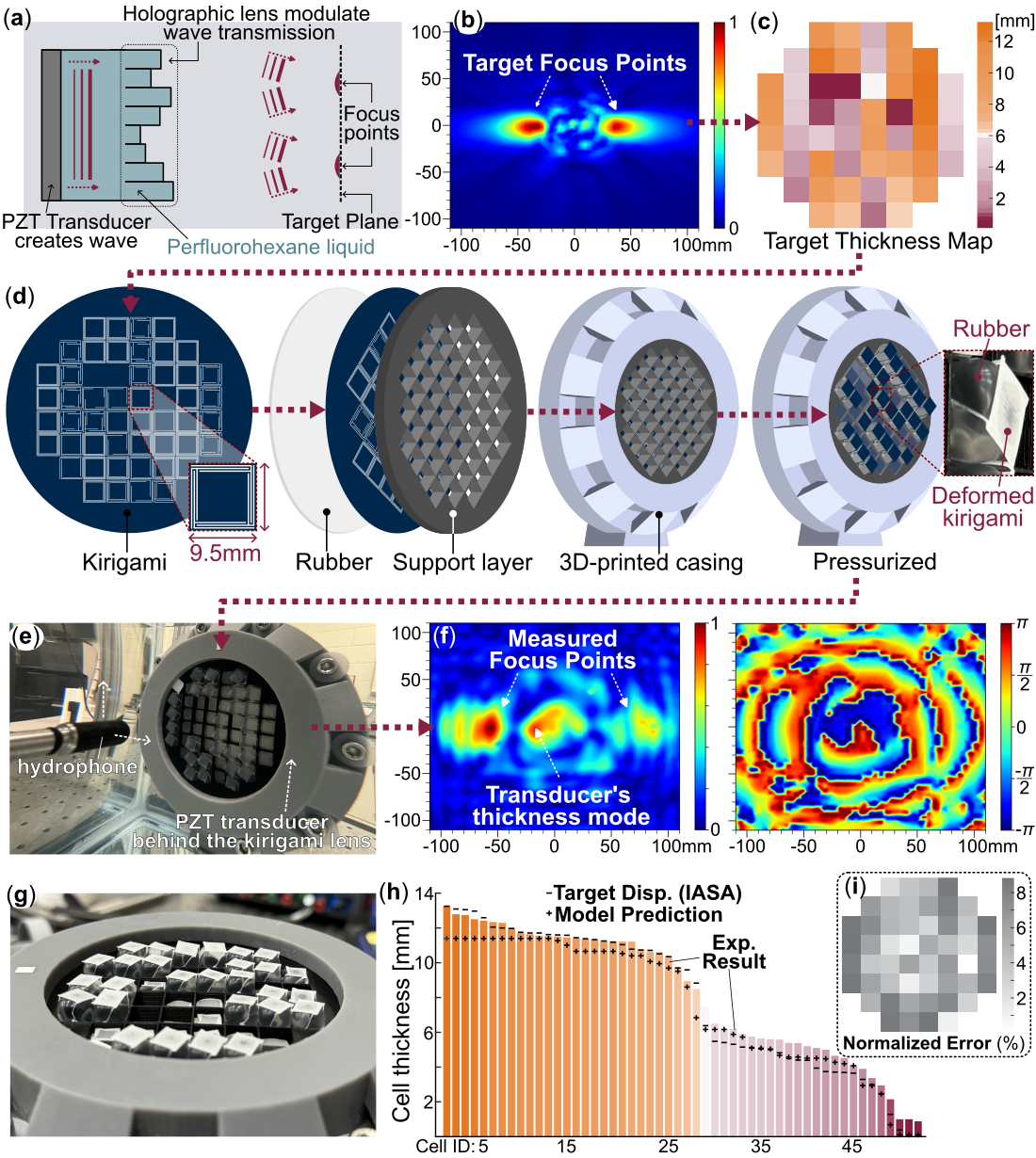}
    \caption{\textbf{The making and testing of a fluidic Kirigami holographic lens for ultrasonic waves.} 
    (a) A schematic diagram showing the fundamental working principle of an acoustic holographic lens. Notice that an accurate thickness distribution of the holographic lens is crucial for steering the acoustic waves. 
    (b) The targeted two focus points in this demonstration. 
    (c) The desired lens thickness map created by the IASA method, involving 52 cells of 9.5mm$\times$9.5mm size. 
    (d) From left to right: The Kirigami sheet is designed by matching the target lens thickness map from IASA and the design space in Fig.~\ref{fig:Scut}e $\rightarrow$ then this Kirigami is sandwiched between a 1mm thin silicone rubber sheet and 6mm thin 3D-printed PLA support layer $\rightarrow$ then everything is encapsulated in 3D-resin-printed casing filled with working liquid inside (perfluorohexane) $\rightarrow$ finally, upon pressurization, the Kirigami cells will morph into a holographic lens with a correct thickness map. 
    (e) The fluidic Kirigami holographic lens under testing. The whole setup is submerged underwater.
    (f) The measured acoustic pressure at the target plane and the corresponding phase distribution. The two focus points are evident. 
    (g) A close-up of the fluidic Kirigami ultrasonic lens under pressure. 
    (h) Comparing the actual thickness map with the targeted values. For visual clarity, we re-arrange the 52 cells according to their out-of-plane thickness. The bars are measured unit cell thickness using an optical sensor (optoNCDT 2300-50 laser triangulation sensors, Micro-Epsilon) , ``$-$'' denote the targeted thickness from the IASA method, and ``$+$'' are the analytical model predictions. 
    (i) The corresponding thickness error map as a percentage of acoustic wave length at 70kHz.}
    \label{fig:holo}
\end{figure}

\subsection*{Demonstrating high-precision shape morphing with ultrasonic holographic lensing} 
To exemplify these two advantages of fluidic Kirigami metasurface for high-precision shape morphing, we apply this concept and create a programmable ultrasonic holographic lens (Fig. \ref{fig:holo}). Acoustic holography is a technique to reconstruct a three-dimensional sound field and visualize sound propagation by leveraging wave interference~\cite{melde2016holograms,mueller1971acoustic,veronesi1987nearfield,hildebrand2012introduction, sallam2024multi}. Such holography can be achieved by specialized lenses, which have a carefully defined and non-smooth thickness map. As the sound wave front travels through or reflects from the lens, this complex distribution of thickness will modulate the wave front's phase and amplitude~\cite{brown2019phase}. As a result, the acoustic wave energy will be focused into pre-defined patterns with diffraction-limited resolution at the target plane \cite{melde2016holograms} (Fig.~\ref{fig:holo}a). Typically, a holographic lens is entirely static (e.g., a 3D-printed disc ~\cite{castineira2020cylindrical, maimbourg20183d, sallam2024gradient}) or entirely active (e.g., phased array transducers (PATs) ~\cite{tseng2017phased,xie2016acoustic,zhao2014manipulation}). The static ones are limited to one particular holographic pattern, while the active ones come with a high cost, especially with high-frequency applications ~\cite{shung2009high,lockwood1996beyond,silverman2016focused,darmani2022non}. The fluidic Kirigami holographic lens can strike a balance between cost and versatility.

As a demonstration, we designed a fluidic Kirigami holographic lens that can steer plane acoustic wavefronts and focus them into two distant points at the target plane (Fig.~\ref{fig:holo}b). To design such a lens, one can integrate the parametric design space as shown in Fig.~\ref{fig:Scut}e and the Iterative Angular Spectrum Approach (IASA) ~\cite{mellin2001limits,shen2019p, xu2023programmable}. The IASA typically starts from an initial holographic lens design and iteratively updates this design by repeating five consecutive steps. (1) Propagating the acoustic field from the initial lens to the target plane---using numeric simulation---and evaluating the quality of the projected acoustic ``image''. (2) Imposing the desired amplitude pattern on the target plane (e.g., the two focus points in our example) while retaining the phase. (3) Propagating the adjusted acoustic field from the target plane back to the hologram plane. (4) Computing the lens thickness map. (5) Updating the complex amplitude at the hologram plane accordingly and repeating the steps.

%the pressure field can be easily transformed from the natural pressure distribution of the acoustic source to the desired pattern through the activation of the design. 

We first conducted a characterization experiment to obtain the source complex amplitude without the Kirigami sheet (a flat surface). This result was then treated as the ``initial lens design'' for obtaining the desired thickness map using IASA. We assume a 1D propagation within the structure, neglecting shear waves and attenuation. The transmission losses are considered minimal and, therefore, also neglected.
Fig.~\ref{fig:holo}b presents the simulated wave field amplitude at the target plane, and Fig. \ref{fig:holo}c shows the final thickness map targets from the IASA.
Then, we can simply find the corresponding Kirigami cut design by matching the target thickness with the parametric design space in Fig.~\ref{fig:Scut}e.
It is noted that, with the available number of Kirigami cells (52 in our case), the image accuracy is not optimal. The pressure at the center, generated by the piezoelectric disk, is not entirely eliminated and directed toward the two points. This issue could be mitigated by increasing the operating frequency, thereby increasing the total number of cells. However, designing smaller cells while maintaining similar levels of out-of-plane thickness poses a significant fabrication challenge and is beyond the scope of this work. Section 3 of the supplement materials derive the ultrasonic frequency limit using current fabrication setup.

Fig.\ref{fig:holo}e-f shows the final fluidic Kirigami lens deformation under global pressure and the experimentally measured pressure amplitude and phase distribution measured by the hydrophone (Section 4 of the supplement information details the experimental setup). The activated state of the Kirigami cells leads to the formation of a hologram-like metasurface (Fig.~\ref{fig:holo}e), which introduces a phase offset for the transmitted acoustic wave to achieve the desired pressure fields --- a two-point pattern in this case (Fig.~\ref{fig:holo}f). 
It is worth highlighting that the displacement of each Kirigami cell agrees well with the targeted thickness map despite the additional rubber sheet, which is not modeled in our theory (Fig.~\ref{fig:holo}e-g). Since this setup functions as a holographic metasurface, the associated error in cell displacements must not exceed the sub-wavelength threshold (\(\lambda/10\)) of the surrounding medium (water). This corresponds to an allowable error of less than 2.11 mm at 70 kHz for each Kirigami cell. The fluidic Kirigami metasurface successfully reaches this tight tolerance with \(\leq 2 \, \text{mm}\) error, indicating its capability to achieve high-precision morphing despite the additional rubber sheet beneath. 

However, despite the very accurate shape morphing from the Kirigami metasurface, the pressure fields observed in the experiments do not perfectly align with the simulation data (i.e.~comparing Figs.~\ref{fig:holo}b and \ref{fig:holo}f). 
%The physical phase distribution at the metasurface plane used to manipulate the wavefront generated during the experiments may differ slightly from the desired thickness map for a two-point pattern. 
This discrepancy may be attributed to several factors, including the effect of gravity, differences in hydrostatic pressure on the Kirigami cells due to the interaction of the two fluids (water and perfluorohexane) at their surfaces, and the bulging of the rubber sheet through Kirigami sheets. Most notably, the IASA simulation assumes that each cell has a perfect cuboid shape with sharp edges, but the inflated fluidic Kirigami cells have a more dome-like shape. Such differences in cell shape can lead to inaccuracies in directing the acoustic energy to the two focus points and may result in the unintended preservation of the natural thickness mode of the transducer, which appears as an additional point at the center between the two focal dots (Fig.~\ref{fig:holo}f). 
%
%In contrast, the deflated state of the Kirigami lens does not generate a phase offset, resulting in a flat surface that transmits only the natural mode of the transducer at the specified working frequency. This condition effectively creates a single focal point at the center, as illustrated in the supplementary material (see Fig.\ref{figure_reference}). 
%
Increasing the working frequency to the lower megahertz scale is the most effective strategy to replicate the two-point focus pattern accurately. However, this would require new micro-fabrication techniques, as we discussed earlier, and will be the topic of future work. 

% However, this adjustment poses challenges, as it would also reduce the aspect ratio of the Kirigami cells, complicating their fabrication due to the smaller dimensions required. Recognizing these challenges, our future studies will rigorously investigate these limitations. By addressing these fabrication hurdles, we aim to enhance the resolution of the Kirigami-based acoustic lens, thereby enabling its effective operation at higher frequencies and unlocking new possibilities for advanced acoustic applications.

\section*{Circular-cut Kirigami Surface for High-Resolution Force Control}
When viewed from the top, a circular-cut Kirigami cell consists of multiple concentric ``rings'', each containing several thin and interconnected curved ligaments (or beams).  Fig.~\ref{fig:Ccut}a-d highlights one such curved ligament, which connects to other ligaments \textit{in the same layer} at its ends and connects to the ligaments \textit{in the inner layer} at its center. 

\subsection*{Analytical tool for analyzing the circular cut Kirigami cells' deformation}
Unfortunately, analyzing the out-of-plane deformation of these curved structures becomes much more complicated than the simple straight ligaments in the square-cut Kirigami; solving the nonlinear Euler-Bernoulli equation on a curved beam becomes mathematically intractable.  To this end, we observe that --- as the circular Kirigami cells undergo out-of-plane deformation --- the outer rings of curved ligaments deform more significantly than the inner rings because the former have slender shapes.  As a result, the outer rings will reach plastic yielding first.
Therefore, \emph{we proposed a novel approach by approximating the large amplitude and nonlinear out-of-plane deformation of the circular-cut Kirigami with a piecewise linear model}, so that: 
\begin{equation}
    \Delta = \sum  \delta^{(i)}, \text{ where } \delta^{(i)} = 
    \begin{cases} 
    \delta^{(i)}(p_0), \text{ before elastic limit }  (S^{(i)}_V\leq S_{yield}),\\
    \delta^{(i)}_\text{max}, \text{ beyond elastic limit } (S^{(i)}_V > S_{yield}).
    \end{cases}\label{eq:circle_disp}
\end{equation}

\noindent Here, $p_0$ is the pressure input. $\delta^{(i)}(p_0)$ is the curved beam's vertical deformation before reaching the elastic limit, correlating linearly to $p_0$ (again, the superscript $^{(i)}$ explains in which ring this curved ligament is located). Meanwhile, $\delta^{(i)}_\text{max}$ is a constant value corresponding to the ligament deformation beyond the elastic limit.  $S^{(i)}_V$ and $S_{yield}$ is the Von Mises stress and yield stress, respectively. As the circular-cut unit cell deforms under pressure, its outer rings of thin beams deform more significantly than the inner layers because the former have more slender shapes.  As a result, these outer layers will reach an ``elastic limit'' first (more on this later), creating a nonlinear behavior even though the beam deformation follows linear mechanics. In what follows, we detail these two deformation stages for a curved beam (while dropping the superscript $^{(i)}$ for clarity).

\begin{figure} [t]
    \centering
    \includegraphics[scale=1.0]{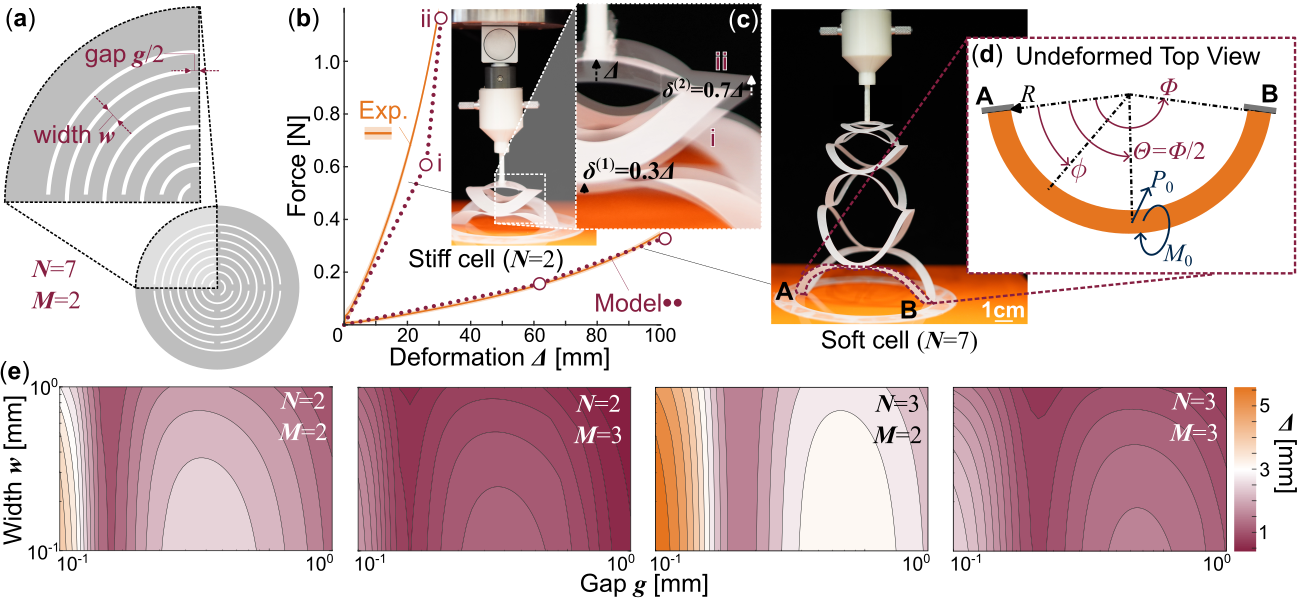}
    \caption{\textbf{The anatomy and design space of circular-cut Kirigami} 
    (a) An example design of a circular-cut Kirigami cell consisting of 7 concentric rings of cuts ($N=7$), with each ring having 2 interconnected ligaments ($M=2$). Again, we label the outermost ring by superscript $^{(1)}$.
    (b) The out-of-plane force-deformation relationships of soft and stiff circular Kirigami cells. Notice that the model predictions (maroon dotted line) are only piecewise linear, but they agree reasonably well with experiment data (orange solid line). 
    (c) A composite image comparing the Kirigami deformation at two different deformation stages. At the high deformation range, the inner ring deforms much more significantly than the outer ring.  This is consistent with the piecewise linear model assumption. 
    (d) Kinematics of an undeformed ligament.
    (e) The modeled out-of-plane deformation corresponding to different circular Kirigami designs.
    }
    \label{fig:Ccut}
\end{figure}

\underline{Before elastic limit}: According to \textit{Castigliano’s second theorem}, the deformation of a beam is governed by:
\begin{equation} \label{eq:1}
    \delta =\frac{\partial{U}}{\partial{Q}},
\end{equation}
where \( U \) is the strain energy and \( Q \) and \( \delta \) are the generalized force and displacement, correspondingly. While the curved beam deformation in the Kirigami cell can be complex, careful observation of the experiment results suggests that it predominantly bends out-of-plane without significant twisting, shear, or stretching deformations (Fig.~\ref{fig:Ccut}b,c).  In addition, the geometry and loading of the Kirigami cell are axisymmetric.  Therefore, we can assume each curved beam is fixed at both ends and subjected to two external loads at its center: an out-of-plane force $P$ and a twisting moment (torque) $T$ in a plane normal the beam's curved axis.

Therefore, denote $\phi$ as the angular position along the curved beam's span, and $\Phi$ is the beam arc's total span ($\phi \in [0\ldots\Phi]$) (Fig.~\ref{fig:Ccut}d). The strain energy is given by \cite{young2002roark}:

\begin{equation} \label{eq:2}
    U \simeq U_f = \int_{0}^{l} \frac{M(\phi)^2}{2EI} \, ds = \int_{0}^{\Phi} \frac{M(\phi)^2}{2EI} R \, d\phi,
\end{equation}
where \( U_f \) is the complementary flexure energy, \( EI \) is the flexural rigidity along the bending axis, \( R \) is the curved beam's radius, and \( l \) is the beam length. Since there are two external loads ($P$ and $T$), the reaction bending moment \(M(\phi) \) at position $\phi$ consists of two components in that \cite{young2002roark}:
\begin{equation} \label{eq:moment}
    \begin{aligned}
        M(\phi)_{P} &= V_{A,P} R \sin \phi + M_{A,P} \cos\phi - T_{A,P} \sin \phi - P R \sin (\phi - \Theta) \langle \phi - \Theta \rangle^0 ,\\
        M(\phi)_{T} &= V_{A,T} R \sin \phi+M_{A,T} \cos \phi-T_{A,T} \sin \phi-T \sin (\phi-\Theta)\langle \phi-\Theta\rangle^0,
    \end{aligned}
\end{equation}
where \( V_A \), \( M_A \), and \( T_A \) are the reaction force, bending moment, and twisting moment at the curved beam's left end $A$, respectively.  $\Theta$ is the external loads' position, and in this case, $\Theta=\Phi/2$. The angle brackets \(\langle \phi-\Theta \rangle^n\) are defined such that \(\langle \phi-\Theta \rangle^n = 0\) when $ \phi<\Theta$, and \(\langle  \phi-\Theta \rangle^n = (\phi-\Theta)^n\) otherwise. 
Based on the formulations above, the work of \cite{young2002roark} showed that the reaction forces and moments at the end of the beam are linearly correlated with the external load in that: 
\begin{equation}
    \begin{aligned}
        V_{A,P} & = \alpha_V P, \; M_{A,P}  = \alpha_M P R, \; T_{A,P}  = \alpha_T P R,\\
        V_{A,T} & = \beta_V \frac{T}{R}, \; M_{A,T}  = \beta_M T, \; T_{A,T}  = \beta_T T,
    \end{aligned}\label{eq:VMT}
\end{equation}
where $\alpha$ and $\beta$ are non-dimensional constants defined by the curved beam's geometry and consecutive properties. Detailed derivations of these constants are available in Section 5 of the Supplement Materials.
The curved beam's out-of-plane deformation becomes:
\begin{equation}
    \delta = \delta_{P} -  \delta_{T}.\label{eq:delta}
\end{equation}
Notice the negative sign indicates that the twisting moment opposes the out-of-plane force. Moreover:
\begin{equation}
\begin{aligned}
   \delta_{P} &= M_{A,P}\frac{R^2 F_1}{EI} + T_{A,P}\frac{R^2 F_2}{EI} + V_{A,P}\frac{ R^3 F_3}{EI} - P\frac{R^3 F_{a3}}{EI},\\
  \delta_{T}&=M_{A,T}\frac{R^2 F_1}{E I}+T_{A,T} \frac{R^2 F_2}{E I} + V_{A,T}\frac{ R^3 F_3}{E I} +T\frac{R^2 F_{a 2}}{E I}.
\end{aligned}\label{eq:l_delta}
\end{equation}

Here, $F_i$ are also non-dimensional constants detailed in Section 5 of the Supplement Materials.
We assume the contact between the pressurized rubber sheet and the Kirigami sheet occurs primarily at the center of its cells, allowing us to approximate the overall out-of-plane force as a concentrated point load at the cell center. Therefore, in the fluidic metasurface, $P=p_0 A / M$, where $p_0$ is the global fluidic pressure, $A$ is the unit cell's area in contact with the Kirigami sheet, and $M$ is the number of curved beams in each layer. The twisting moment is $T=p_0 A \times R/M$. Eqs. (\ref{eq:delta}) and (\ref{eq:l_delta}) give the linear deformation mechanics as described in the first line of Eq. (\ref{eq:circle_disp}).

\underline{Beyond elastic limit}: While Eq. (\ref{eq:l_delta}) describes the linear elastic deformation of the curved beam, we observe that a nonlinearity occurs when the outer rings reach their elastic limit before the inner rings. Here, we define the elastic limit based on yield stress. When the bending stress $S_{M_A}$ reaches the yield stress $S_{yield}$ at the beam's fixed end (where the bending stress is maximum), the elastic limit is reached. 
As shown in Fig.~\ref{fig:Ccut}c, it deforms much slower after the outer layer of curved beams reaches the elastic limit. Meanwhile, the inner layers continue to bend. Therefore, we assume the curved beam deflection is constant after the limit is reached (i.e., the second line of Eq. (\ref{eq:circle_disp})). 
The \textit{Von Mises stress} $S_V$ is:
\begin{equation}
S_V = \sqrt{S_{M_A}^2 + 3\left(S_{T_A} + S_{V_A}\right)^2} \simeq S_{M_A} = \frac{M_A c}{I},
\end{equation}
where $I$ is the area moment of inertia, and $c=t/2$ is the distance between the beam's neutral axis and its surface. Since in the yield region $S_V = S_\text{yield}$, the maximum reaction bending moment is:
\begin{equation}
        M_{A,\text{max}} = \frac{S_\text{yield}I}{c}.
\end{equation}

Based on the tensile test results, \(S_{yield} \approx 160 \, \text{MPa}\). Using the second and fourth equations in Eq. (\ref{eq:VMT}), we can solve for \(P_\text{max}\) and \(T_\text{max}\), representing the external loads at the elastic limit. We then use the rest of Eq. (\ref{eq:VMT}) to solve for the corresponding reaction force and moments $V_{A,P_\text{max}}$, $T_{A,P_\text{max}}$, $V_{A,T_\text{max}},$ and $T_{A,T_\text{max}}$. Finally, we can plug in these maximum reaction forces and moments into Eq. (\ref{eq:delta}, \ref{eq:l_delta}) to obtain the curved beam deformation at the elastic limit $\delta_\text{max}$.  

We tested the accuracy of this piecewise linear model on two unit cell samples (Fig.~\ref{fig:Ccut}b). One cell is relatively stiff with $N=2,$ $M=2$, $d=70$mm, $w=10$mm, and $g=10$mm, while the other cell is softer with $N=7,$ $M=2$, $d=80$mm, $w=5$mm, and $g=5$mm. Certainly, the piecewise linear model is not as accurate as the fully nonlinear model discussed earlier, but it still provides a reasonable prediction without unnecessary mathematical complexity.

Moreover, based on this validated analytical model, one can also explore the circular-cut Kirigami's design space for prescribing its out-of-plane deformation. Fig.~\ref{fig:Ccut}e summarizes the results of such a parametric study where the unit cell size is $d=9.5$mm, the number of concentric rings varies from $N=2$ to $N=3$, and the number of curved ligaments in each ring varies from $M=2$ to $M=3$. The actuation pressure is set at 55kPa. By adjusting the curved beam's width $w$ and gap size $g$, one can continuously prescribe the unit cell deformation from 0 to 6mm (i.e., 0.67 $\Delta/d$ aspect ratio). Moreover, the design mapping is no longer 1-to-1.  There can be multiple design choices for a targeted out-of-plane deformation $\Delta$, giving us the flexibility to consider other design requirements (as we detail in the example below).

%\subsection*{Demonstrating high-resolution force control with a haptic device}
\subsection*{Using fluidic Kirigami for high-resolution force control in a haptic device}

\begin{figure}[h!]
    \centering
    \includegraphics[scale =1]{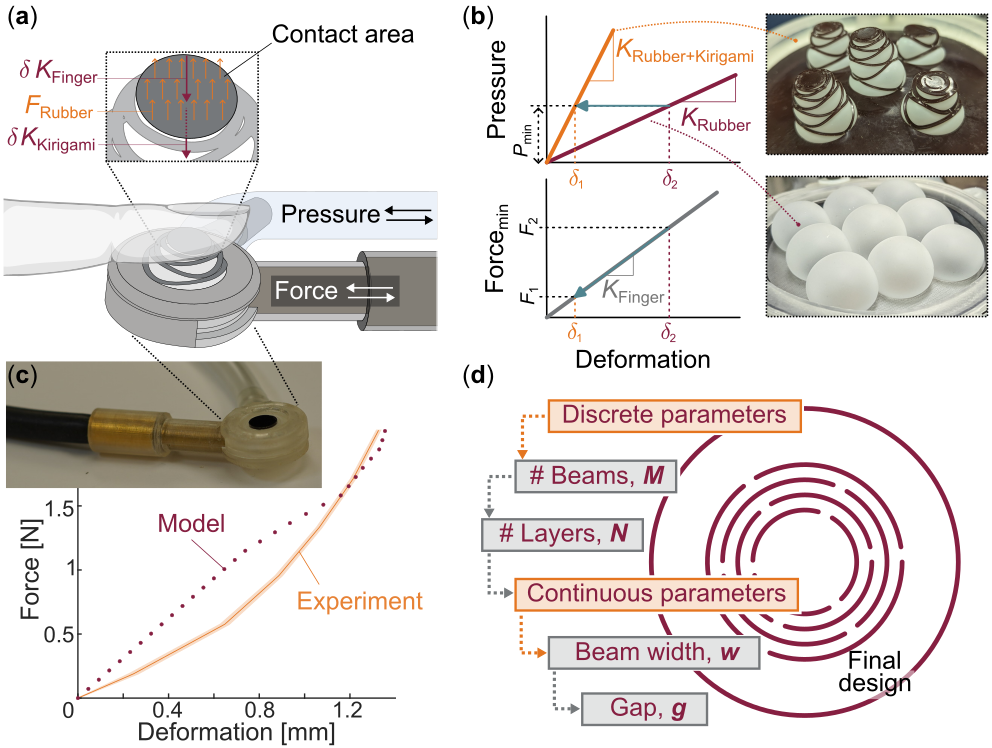}
    \caption{\textbf{Designing fluidic Kirigami for haptic interfacing}
(a) This haptic device intends to apply a precise and low-amplitude force stimulus---via a fixed contact area---to an individual's finger. The magnitude of the force stimulus can be adjusted by modulating the fluidic air pressure using a pneumatic control system. The cut pattern design of the Kirigami layer dictates the force-deformation relationship, which can be programmed with our piecewise linear model. 
(b) Here, we use illustrative force-deformation curves with and without the Kirigami layer to explain the benefit of fluidic Kirigami. When the force stimulator makes contact with the individual's finger, the displacement of the Kirigami and finger are the same (denoted as $\delta_i$). Due to the stiffness increase from Kirigami, the finger displacement decreases from $\delta_{2}$ to $\delta_{1}$ for the minimum pressure increment. As a result, the force transmitted to the finger decreases from $F_2$ to $F_1$, thus providing a higher-resolution stimulus.
(c) The experimentally measured free-stroke force and displacement relationship agrees reasonably well with the theoretical prediction. For the experimental data, a fiber optic sensor tracked the displacement of the Kirigami's contact area, as shown in (a), and a pressure sensor estimated the force applied. The solid line is the averaged data from 5 loading/unloading cycles, and the thin shadow is the corresponding standard deviation, showing highly repeatable performance.
(d) Design flow chart for the Kirigami considering the high-resolution force output: The discrete parameters of the number of beams ($M$) and number of rings ($N$) allow for greater changes in the force-displacement curve, whereas the continuous parameters of the beam width ($w$) and gap ($g$) allow for fine-tuning.}
\label{fig:haptic}
\end{figure}

Prior work on haptic devices that controlled force output with a high-resolution and low magnitude primarily relied on kinematic (rigid) mechanisms or piezoelectric materials~\cite{giraud_haptigami_2022,yeh_application_2020,gurari_customization_2014}. Fluidic Kirigami offers a new potential for creating these haptic devices, with an added benefit of supporting designs that can be highly adaptable and customizable, including in small work areas, depending on the Kirigami cut pattern (Fig.~\ref{fig:haptic})~\cite{iiyoshi_origami-inspired_2024}. Here, we present an example of using the circular-cut Kirigami mechanics model to guide the design of a haptic device that can apply, modify, and measure forces that are small in amplitude (e.g.,~0.01N). This example shows how to achieve a desired elastic force and displacement range by adjusting the stiffness generated by the Kirigami layer.

In this example, we aim to create a compact haptic ``button'' that can deliver forces in the range of $0-4$N and displacements in the range of $0-1.5$mm, with force increments lower than the threshold of human force detection (0.05N) (Fig.~\ref{fig:haptic}a)~\cite{gurari_customization_2014,bretz_force_2010}. One of the key considerations here is force reproducibility~\cite{gassert_actuation_2006}. This haptic device should precisely and repeatedly deliver the same low amplitude force without relying on closed-loop control, which can become problematic when working with delays inherent to pneumatic systems~\cite{turkseven_model-based_2018}. Therefore, the deformation must remain within the elastic region to allow repeatable force application. Our previously discussed piecewise linear model can readily predict the applied force and expected displacement while remaining within the elastic deformation limit.

Our piecewise linear model highlights four key geometrical variables that govern the force-displacement response of a circular-cut Kirigami cell: the number of ``rings'' defined by the concentric cuts $N$, the number of curved beams in such a ring $M$, the width of these curved beams $w$, and the gap between adjacent cut tips $g$ (Fig.~\ref{fig:Ccut}a). In principle, one can achieve the desired force-deformation output by looking up the design space in Fig.~\ref{fig:Ccut}e. However, the order of selecting these design parameters is critical to ensure that the fluidic Kirigami remains within the elastic region. Because stress concentrates near the tips of slit cuts, the number of curved beams in each ring $M$ has the most significant influence on the elastic deformation limit. Additionally, the number of rings $N$ strongly constrains the overall magnitude of the elastic deformation range. Therefore, we first selected $M$, followed by $N$, using the design space in Fig.~\ref{fig:Ccut}e. Next, we selected the curved beam width $w$, followed by the cut gap $g$ to fine-tune the fluidic Kirigami's mechanical response. Unlike $M$ and $N$, the final two variables are continuous, offering considerable design freedom. Fig.~\ref{fig:haptic}d summarizes the overall design flow chart.

For the example haptic interface shown here, we chose three beams in each ring ($M = 3$) to ensure an elastic response within the desired force/deformation range (Fig.~\ref{fig:haptic}c). The number of rings, $N$, was set at 3 to guarantee sufficient deformation while ensuring easy fabrication of the Kirigami layer. The gap and beam width were set at $0.7$mm and $0.7$mm, respectively, to achieve the targeted $0-4$N output force. Fig.~\ref{fig:haptic}c shows the final prototype based on the aforementioned fluidic Kirigami design (more haptic experiment details are available in Section 6 of the supplement materials). The experimentally measured output force and displacement match the theoretical predictions, with both remaining within the elastic region. Therefore, this Kirigami layer allows for repeatable deformation when the input pressure is low, achieving the modeled force-displacement curve. 

An added benefit of including the Kirigami layer is that the increased stiffness when deforming out-of-plane permits higher resolution in force control. This benefit arises particularly when the stiffness of the rubber sheet is small in comparison to the combined Kirigami and rubber layer. As shown in Fig.~\ref{fig:haptic}b, the stiffness of the combined rubber and Kirigami layer is greater than of the rubber layer alone. For a minimum change in the input pressure (due to hardware limitations), there is a minimal change in force that can be applied $F_{min}$. This minimal pressure-generated force $F_{min}$ results in a smaller out-of-plane deformation (i.e.,~displacement) for the combined rubber and Kirigami layer than for the rubber sheet alone. When interacting with an object with a stiffness (e.g.,~finger), the force applied to the object will be smaller for the combined rubber and Kirigami layer than the rubber layer alone since the combined rubber and Kirigami layer will generate a smaller displacement $\delta$ (i.e.,~$F_{\mathrm{StiffObject}}=k_{\mathrm{StiffObject}}\cdot\delta$). Hence, by modifying the stiffness of the Kirigami layer, this haptic device enables the desired resolution in the applied force. 
Another benefit of the Kirigami layer is its safety and functionality as a haptic device. When inflating with a large pressure, the hardware (e.g., the rubber sheet) will fail due to bursting and/or breakdown. The additional stiffness from the Kirigami layer will increase the failure pressure, thereby increasing the maximum usable force.

While we focus primarily on the force control capabilities, we also would like to highlight the benefits of the Kirigami layer for sensing. The cut-free center region provides a mounting surface for a reflecting film used with a sensor (e.g.,~fiber optic sensor), allowing accurate displacement measurement and better force estimation. This accurate measurement would not be possible if only the rubber sheet was included, since inflation could occur in varying directions. 

% Finally, we note that an additional consideration in the selection of the cut pattern is whether to induce shear forces in addition to the normal force. If desired, a cut design could incorporate multiple layers intentionally misaligned to induce rotation, thereby stretching the skin. This behavior was not of interest for our haptic interface, so the selected cut pattern was designed to allow only vertical movement.

\section*{Discussion}
In this study, we developed a fluidic Kirigami metasurface capable of precise shape morphing and control of the output force, leveraging geometric design principles rather than complex material engineering. Through theoretical modeling and experimental validation, we demonstrated how square and circular-cut Kirigami cells enable tunable morphing and how to apply them to different multiphysics applications like ultrasonic holography (high-precision out-of-plane deformation) and haptic interfaces (high-resolution out-of-plane force control). Our results show that fluidic kirigami metasurface provides a scalable, easily manufacturable solution for morphing surfaces, as cutting is an inherently accurate and scalable fabrication method.

Another notable contribution of this study is the new nonlinear mechanics model of the square and circular-cut kirigami cell. We adopted two approaches to analyze the Kirigami deformation according to their design: A fully nonlinear Euler beam approach for the straight ligaments in the square-cut Kirigami and a piecewise-linear approach for the curved ligaments in the circular-cut Kirigami. Either of these two approaches (or a combination of them) could apply to other Kirigami designs.

Regarding future work, the fluidic Kirigami metasurface concept could benefit from more advanced micro-fabrication methods. This would enable smaller and denser unit cells for new functions (e.g., pushing the usable frequency of ultrasonic holographic lensing to MHz). Including rubber sheets in the mechanics, model will further improve modal accuracy. Finally, directly embedding the Kirigami layer inside the rubber sheet and eliminating the friction contact between them could allow for a more consistent output at higher pressure. Finally, by refining the control and scalability of these morphing surfaces, this study lays the foundation for next-generation adaptive materials with broad applications in acoustic manipulation, haptic feedback, and reconfigurable structures.

\section*{Acknowledgements}

A. Kahak and S. Li acknowledge the partial support from the National Science Foundation (CMMI-2312422), the Department of Energy (InDEEP Prize), and Virginia Tech (via startup fund).  N. Gurari acknowledges the partial support from the National Institutes of Health (K25HD096116), Virginia Tech Institute for Critical Technology and Applied Science (Junior Faculty Program), and Virginia Tech (via startup fund). M. Sayed Ahmed, H. Kulkarni and S. Shahab acknowledge the partial support from the U.S. National Science Foundation (NSF) under CAREER grant Award No. CMMI 2143788.

\section*{Author contributions statement}
S.L., S.S., and N.G. conceived the studies. A.K. derived and experimentally validated the mechanics model. M.A. and H.K. conducted the acoustic experiments. N.K. conducted the haptic experiments. All authors drafted and edited the manuscript.

\section*{Additional information}
\textbf{Competing interests:} The authors declare no conflict of interests. 

\bibliography{sample}

\newpage

\section*{Supplement Section 1, Generic Solution of Nonlinear and Straight Cantilever Beam}
If the deformed nonlinear cantilever beam has a deflection point along its beam span, one can assume the external bending moment $M_0$ is negative, the transverse force $P_0$ is positive, and $0 \leq M_0 \leq \sqrt{2P_0EI}$. Moreover, the generic solution to Eq. (\ref{Non-linear}) in the main text --- without applying any explicit assumption on beam tip rotation $\theta_0$ --- is \cite{kimball2002modeling}:
\begin{equation} \label{elliptic equations}
\begin{aligned}
\alpha &= f^* ,\\
\beta_x &= \frac{x_0}{l} = \frac{\sqrt{2}}{\alpha} c_x^* ,\\
\beta_y &= \frac{y_0}{l} = \frac{1}{\alpha} \left(f^* - 2 e^* + \sqrt{2} c_y^* \right),
\end{aligned}
\end{equation}
where \( f^* \), \( e^* \), $c_x^*$  , and $c_y^*$ are functions defined as:
\begin{equation}
    \begin{aligned}
    f^*&=F\left(\gamma_1, k\right)+F\left(\gamma_2, k\right), \\
    e^*&=E\left(\gamma_1, k\right)+E\left(\gamma_2, k\right), \\
    c_x^*&=c_{x 1}+c_{x 2}, \\
    c_y^*&=c_{y 1}+c_{y 2}.
    \end{aligned}
\end{equation}

Here, $F(\zeta, \omega)$ and $E(\zeta, \omega)$ are the complete elliptic integral of the first and second kind, respectively:
\begin{equation}
    \begin{aligned}
 F(\gamma, \kappa) & \equiv \int_0^{\gamma} \frac{1}{\sqrt{1-\kappa^2 \sin ^2 \nu}} d \nu, \\
 E(\gamma, \kappa) & \equiv \int_0^{\gamma} \sqrt{1-\kappa^2 \sin ^2 \nu} d \nu,
\end{aligned}
\end{equation}
and
\begin{equation}
    \begin{aligned}
        c_{x 1} &= c_{y1} = \sqrt{\lambda}, \\
c_{x 2} &=\sqrt{\kappa}, \\
c_{y 2}&=\sqrt{\kappa}\left(\sqrt{\frac{1+\sin \theta_0}{1-\sin \theta_0}}\right), \\
\gamma_1  &=\sin ^{-1} \sqrt{\frac{2 \lambda}{\lambda+1}}, \\
\gamma_2 &= \sin ^{-1} \sqrt{\frac{2}{\lambda+1}\left(\frac{\kappa}{1-\sin \theta_0}\right)}, \\
k &=\sqrt{\frac{\lambda+1}{2 }} .
    \end{aligned}
\end{equation}

If we assume beam tip rotation $\theta_0\approx0$, the solutions above will simplify to what is presented in the main text.

%\newpage
%\section*{Supplement Section 2, Measuring the Elastic Modulus of Plastic Sheets}

\newpage
\section*{Supplement Section 2, Fabrication and Setup of Fluidic Kirigami Morphing Surface}

Activating a fluidic Kirigami morphing metasurface requires a leak-proof chamber with an open side, a support lid layer, and a morphing layer consisting of a rubber sheet and a plastic Kirigami sheet.

For the Kirigami morphing surface shown in Fig.~\ref{fig:vision}b and \ref{fig:Scut}d, commercial T-slot aluminum profiles were used to form a simple chamber, which was mounted on an isolation table (Newport SmartTable) to create the fluid container. Aerovac sealant tape was applied to ensure the container is pressure-proof.

To fabricate the deformable rubber sheets, we used the Ecoflex™ 00-10 rubber (a platinum-catalyzed silicone from Smooth-On Inc.) and an Elcometer 3580 film applicator to produce submillimeter to millimeter-thin rubber sheets. For the Kirigami, we used a low-cost gray plastic shim sheet (10in$\times$20in, 0.0075in thin, from McMaster-Carr) and cut them using a Trotec laser machine (Speedy 360). Any commercial software, such as Inkscape in this study, can be used to generate the cut pattern as a DXF file, which is compatible with the laser cutter.

A 6mm water-jet cut aluminum plate was used as the support lid layer. It has $10\times10$mm square cutouts to fix the perimeters of each Kirigami unit cell. Once all layers are in place, it is pressured with a pressure regulator (Proportion-Air QB1XANEEN14.7P30PSG). It is worth noting that the order of assembly is important: the rubber sheet should be below the Kirigami sheet. All edges of these two sheets should be clamped by the support lid layer and the container's main body. 

For the acoustic holography experiment, we adopt a similar assembly approach except for three differences: the fluid container is 3D-printed resin, the resin lid is separate from the PLA support layer, and the working fluid is Perfluorohexane liquid.

\begin{figure}[h]
    \centering
    \includegraphics[width=1\linewidth]{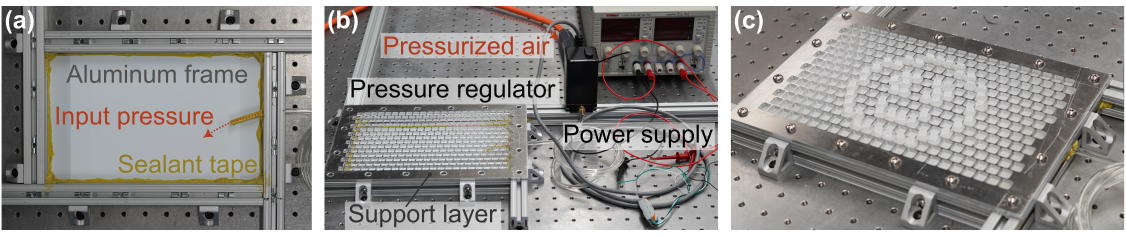}
    \caption{\textbf{Experimental setup of fluidic Kirigami morphing metasurface.} (a) Half-assembled fluidic Kirigami morphing surface system demonstrating its components. (b) A picture of the experimental setup. (c) Smiley face Kirigami morphing surface in pressurized state.}
    \label{fig:fab}
\end{figure}

\newpage
\section*{Supplement Section 3: Analyzing the Working Frequency of Fluidic Kirigami Holographic Lens Based on the Fabrication Limits}

In this study, we define the ``aspect ratio ($AR$)'' of a square-cut Kirigami cell as the ratio between its out-of-plane deformation under pressure over its side length (i.e.,~$\Delta/d$). This aspect ratio is crucial in designing the acoustic hologram because it directly dictates the workable acoustic frequency. Typically, working with higher acoustic frequencies is desirable because it enables higher resolution pressure patterns ~\cite{chen2018deep,baac2012carbon} with a broader range of applications ~\cite{neubach2009ultrasound,kulkarni20244d,tachibana2001use,pahk2015ultrasonic}. However, high working frequencies require a small cell size with a high aspect ratio. Therefore, we must find the maximum acoustic frequency that can be handled by the fluidic Kirigami metasurface lens based on its deformation kinematics and fabrication limits.

For acoustic frequencies below 2 kHz, the center-to-center distance between adjacent Kirigami cells should be equal to or less than half of the acoustic wavelength ~\cite{hansen2001fundamentals,li2013harvesting}, which dictates the unit cell size $d$. This relationship is described by the equation:
\begin{equation} \label{unit cell side length}
\begin{aligned}
    \text{Center-to-center distance between adjacent cells} &= \frac{d}{2} + FS_\text{min} \
    \leq \frac{\lambda}{2}.
\end{aligned}
\end{equation}

\begin{figure}
    \centering
    \includegraphics[scale=1.0]{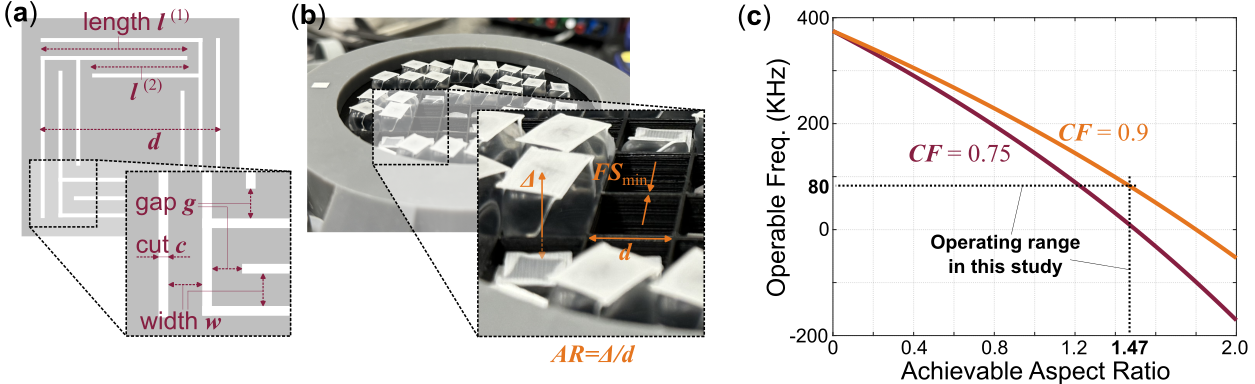}
    \caption{\textbf{The operable acoustic frequency based on the limits of fabrication setup}. (a) The design variable of a square-cut Kirigami unit cell; notice the width of the cut $c$ using laser cutter is non-zero. (b) A zoomed-in view of the pressurized Kirigami holographic lens highlights the definition of aspect ratio ($AR$) and the non-zero wall thickness ($FS_\text{min}$) in the 3D-printed support layer. (c) The correlation between the usable acoustic frequency and achievable aspect ratio, with $c\approx 0.05$mm and $FS_\text{min}\approx 0.4$mm.}
    \label{fig:freq}
\end{figure}

\noindent Here, $FS_\text{min}\approx0.4$mm, and it is the minimal feature size that the 3D printer can print (Ultimaker S5), corresponding to the wall thickness of the 3D-printed support layer (Fig.~\ref{fig:freq}b). 

Suppose all ligaments in a Kirigami cell are deformed to what is \emph{kinematically} possible; they should be stretched into straight linkages. The Kirigami cut has two concentric rings of straight ligaments ($N=2$), so the length of these ligaments $l^{(i)}$ are:
\begin{equation}
\begin{aligned}
    l^{(1)} & = d-w^{(1)}-g^{(1)}-2c ,\\
    l^{(2)} & = d - 2w^{(1)} - g^{(1)} - w^{(2)} - g^{(2)} - 5c,
\end{aligned}
\end{equation}
where the superscript $^{(1)}$ and $^{(2)}$ denote the variables in the outer ring and inner ring, respectively. $w$ and $g$ are the beam width and gap size as defined in Figs.~\ref{fig:Scut}a and \ref{fig:freq}a. $c$ is the width of the actual cut made by the laser cutters or plotter cutters. In this study, we use a laser cutter to cut the plastic Kirigami sheet and $c\approx0.05$mm (Trotec laser machine Speedy 360,  power=\%23.0, velocity=3.0m/s, and frequency=2400Hz).

Therefore, the kinematically maximum aspect ratio of a Kirigami unit cell---without considering the bending stiffness of the plastic Kirigami ligament or stretching stiffness of the pressure-sealing rubber sheet---would be:
\begin{equation} \label{aspect ratio-1}
\begin{aligned}
    AR_\text{max} = \frac{l^{(1)}+l^{(2)}}{d}.
\end{aligned}
\end{equation}

\noindent However, the aspect ratio in realistic testing never reaches this kinematic upper limit. In our experiments, the actual Kirigami cell deformation can reach 75\% to 90\% of this limit before damage occurs. Therefore, we introduce a correction factor $CF=0.75-0.9$ and apply it to Eq. (\ref{aspect ratio-1}) to estimate the achievable aspect ratio based on the current fabrication limits:

\begin{equation} \label{aspect ratio-max}
\begin{aligned}
     AR^* = CF \left(2-\frac{7\left(FS_\text{min}+c\right)}{d}\right).
\end{aligned}
\end{equation}

By integrating the equations above and reinforcing the constraints from our fabrication techniques (i.e.,~3D-printed support layer and laser-cut Kirigami sheets), we can obtain the correlation between the operable range of acoustic frequency and achievable aspect ratio of Kirigami unit cells so that:
\begin{equation} \label{frequency & design}
\begin{aligned}
    \Omega_\text{max} = \frac{0.5v_w}{\frac{7\left(FS_\text{min}+c\right)}{2-AR^{}/CF^{}}+FS_\text{min}},
\end{aligned}
\end{equation}
where $v_w$ is the speed of sound in water at room temperature. The prediction of this equation is shown in Fig.~\ref{fig:freq}c. One can conclude that---based on the current fabrication setup---operating at a higher frequency would reduce the achievable aspect ratio in the Kirigami unit cells, but this trend directly conflicts with the requirement of holographic lensing because it desires a high aspect ratio. The targeted holographic lens thickness map based on the 2-point focus point calls for roughly a 1.5 aspect ratio (Fig.~\ref{fig:holo}c); therefore, we can operate up to the 80kHz range. 

\newpage
\phantom{bla}
\newpage

\section*{Supplement Section 4, Experimental Setup of the Acoustic Holography using Fluidic Kirigami Metasurface}

\begin{figure}
    \centering
    \includegraphics[width=1\linewidth]{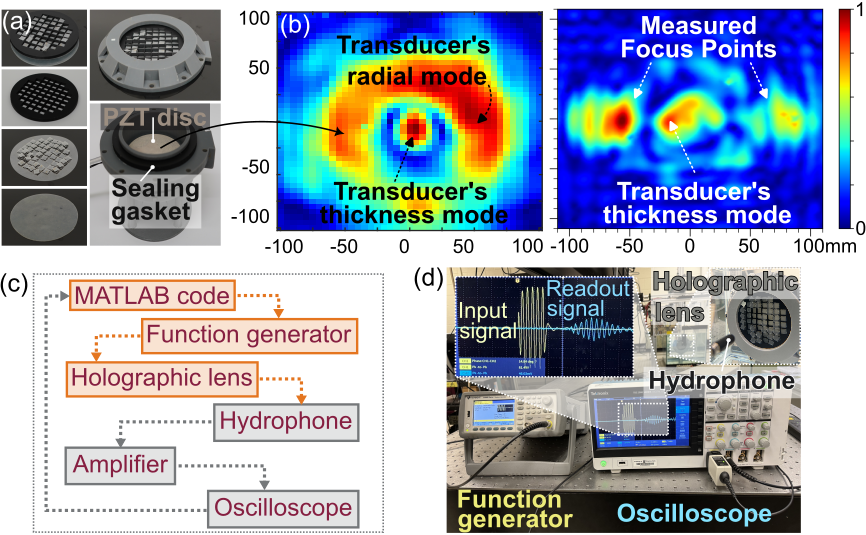}
    \caption{\textbf{Experimental Setup of acoustic holography} (a) Half-assembled acoustic lens demonstrating its components. (b) The measured acoustic pressure at the target
plane for unpressurized and pressurized states. Transducer's thickness mode is visible at the center of the acoustic image for both states which is hard to steer away at low working frequencies.  (c) A schematic diagram of the experimental setup. (d) A signal generator used in this experiment, highlighting the input and readout signals.}
    \label{fig:acoustic_experiment}
\end{figure}

The experiment used a piezoelectric crystal disc (Steminc SM111, 85mm diameter and 6mm thickness). The resonant frequency of this particular disc is 350kHz in the thickness mode, with a resonant impedance of \( \leq 0.49 \, \Omega \). However, the working frequency of this experiment is set at 70kHz so that the Kirigami unit cell can achieve the targeted thickness (or aspect ratio) based on the IASA simulation (as we discuss in detail in Section 3 of the Supplement Materials). The density of the PZT disc is 7600kg/m$^3$, and the sound speed in the PZT material is 3104m/s. The disc is securely held within the acoustic lens by a 3D-printed casing filled with the working fluid perfluorohexane. The speed of sound in this fluid is 600m/s, as determined through a simple transducer-receiver experiment. The entire assembly --- including the PZT disc, the working fluid, the sealing rubber sheet, the Kirigami sheet, and the 3D-printed support layer --- is encased within a 3D-printed cover (Fig.~\ref{fig:acoustic_experiment}a). This setup creates a pressure seal that keeps the components isolated from the surrounding medium (deionized water). Finally, the displacement of the Kirigami cells is manually controlled by the pressure-sealed working fluid using a high-precision syringe pump connected to the acoustic lens. 

The input voltage to the PZT transducer is generated, controlled, and acquired using a waveform function generator (Keysight 33500B series), an amplifier (E\&I RF power amplifier, model A075), and a digital oscilloscope (Tektronix TBS 2000 series, model TBS2104), respectively. The pressure wave is generated by transmitting a burst signal of 5 cycles, with a burst period of 50ms, to the PZT transducer (Fig.~\ref{fig:acoustic_experiment}d). The target plane for focusing the acoustic waves is situated in the water medium, precisely 10cm from the face of the acoustic lens. To measure the acoustic response of the lens at the target plane, we recorded the sound pressure using a calibrated hydrophone (Teledyne Reson TC4013, sensitivity of 130 dB re 1 \textmu Pa/V at 1m at 70 kHz) mounted on a custom-made 3D-positioning system. The hydrophone preamplifier was connected to a DC coupler with a power supply (Precision Acoustics Ltd). We acquire the hydrophone signals using a built-in MATLAB function for data acquisition and a customized script that performs a 2D scan of the acoustic field in the target plane. The hydrophone data was live-processed using the Fast Fourier Transform (FFT) method with MATLAB interfacing. The measurement step size is set to 4mm to scan an area of 100mm by 100mm. The data obtained from these experiments are post-processed through interpolation to achieve higher-resolution acoustic scans.
%\textcolor{red}{The scans are performed under two conditions: when the Kirigami cells are not activated (depressurized) and when they are activated (pressurized)}. 
The actual out-of-plane displacements of the Kirigami unit cells are measured using a laser scanner (optoNCDT
2300-50 laser triangulation sensors, Micro-Epsilon), which helps bring the actual thickness maps of the Kirigami metasurface to values as close as possible to the simulated thickness targets (Fig.~\ref{fig:holo}h).

\newpage
\section*{Supplement Section 5, Additional Formulations for the Piecewise Linear Mechanics Model for Circular-Cut Kirigami}

Eq. (\ref{eq:VMT}) of the main text describes the reaction force and torque at the curved beam's ends as a result of the external force and bending moment at its center.  From the out-of-plane external force $P$, the reaction force and moment are:
\begin{equation} \label{external force reactions}
\begin{aligned}
V_{A,P} &= P \frac{C_{a3}(C_4 C_8 - C_5 C_7) + C_{a6}(C_2 C_7 - C_1 C_8) + C_{a9}(C_1 C_5 - C_2 C_4)}{C_1(C_5 C_9 - C_6 C_8) + C_4(C_3 C_8 - C_2 C_9) + C_7(C_2 C_6 - C_3 C_5)},\\
M_{A,P} &= PR \frac{C_{a3}(C_5 C_9 - C_6 C_8) + C_{a6}(C_3 C_8 - C_2 C_9) + C_{a9}(C_2 C_6 - C_3 C_5)}{C_1(C_5 C_9 - C_6 C_8) + C_4(C_3 C_8 - C_2 C_9) + C_7(C_2 C_6 - C_3 C_5)},\\
T_{A,P} &= PR \frac{C_{a3}(C_6 C_7 - C_4 C_9) + C_{a6}(C_1 C_9 - C_3 C_7) + C_{a9}(C_3 C_4 - C_1 C_6)}{C_1(C_5 C_9 - C_6 C_8) + C_4(C_3 C_8 - C_2 C_9) + C_7(C_2 C_6 - C_3 C_5)},
\end{aligned}
\end{equation}
and from the external torque $T$:
\begin{equation} \label{external torque reactions}
\begin{aligned}
V_{A,T} & = -\frac{T}{R} \frac{C_{a 2}\left(C_4 C_8-C_5 C_7\right)+C_{a 5}\left(C_2 C_7-C_1 C_8\right)+C_{a 8}\left(C_1 C_5-C_2 C_4\right)}{C_1\left(C_5 C_9-C_6 C_8\right)+C_4\left(C_3 C_8-C_2 C_9\right)+C_7\left(C_2 C_6-C_3 C_5\right)},\\
M_{A,T} &= -T \frac{C_{a 2}\left(C_5 C_9-C_6 C_8\right)+C_{a 5}\left(C_3 C_8-C_2 C_9\right)+C_{a 8}\left(C_2 C_6-C_3 C_5\right)}{C_1\left(C_5 C_9-C_6 C_8\right)+C_4\left(C_3 C_8-C_2 C_9\right)+C_7\left(C_2 C_6-C_3 C_5\right)},\\
T_{A,T} &= -T\frac{C_{a 2}\left(C_6 C_7-C_4 C_9\right)+C_{a 5}\left(C_1 C_9-C_3 C_7\right)+C_{a 8}\left(C_3 C_4-C_1 C_6\right)}{C_1\left(C_5 C_9-C_6 C_8\right)+C_4\left(C_3 C_8-C_2 C_9\right)+C_7\left(C_2 C_6-C_3 C_5\right)}.
\end{aligned}
\end{equation}

\noindent Here, $C_i$ are constants based on constitutive material properties and curved beam geometry so that:
\begin{equation}
\begin{aligned}
C_1 &= C_6 = \frac{1 + \xi}{2} \Phi \sin \Phi - \xi \left(1 - \cos \Phi\right) ,\\
C_2 &= C_9 = \frac{1 + \xi}{2} \left(\Phi \cos \Phi - \sin \Phi\right) ,\\
C_3 &= -\xi (\Phi - \sin \Phi) - \frac{1 + \xi}{2} \left(\Phi \cos \Phi - \sin \Phi\right) ,\\
C_4 &= \frac{1 + \xi}{2} \Phi \cos \Phi + \frac{1 - \xi}{2} \sin \Phi ,\\
C_5 &= C_7 = -\frac{1 + \xi}{2} \Phi \sin \Phi ,\\
C_8 &= \frac{1 - \xi}{2} \sin \Phi - \frac{1 + \xi}{2} \Phi \cos \Phi ,
\end{aligned}
\end{equation}
and $C_{aj}$ are defined as:
\begin{equation}
\begin{aligned}
 C_{a 2} &= C_{a 9} =\frac{1+\xi}{2}\left[(\Phi-\Theta) \cos (\Phi-\Theta)-\sin (\Phi-\Theta)\right],\\
C_{a 3} &=-\xi\left[\Phi-\Theta-\sin (\Phi-\Theta)\right]-C_{a 2} ,\\
C_{a 5}&=-\frac{1+\xi}{2}(\Phi-\Theta) \sin (\Phi-\Theta),\\
C_{a 6} &=\frac{1+\xi}{2}(\Phi-\Theta) \sin (\Phi-\Theta)-\xi\left[1-\cos (\Phi-\Theta)\right] ,\\
C_{a 8} &=\frac{1-\xi}{2} \sin (\Phi-\Theta)-\frac{1+\xi}{2}(\Phi-\Theta) \cos (\Phi-\Theta).
\end{aligned}
\end{equation}

\noindent \(\Phi\) is the arc span angle of the entire curved beam. $\Theta= 0.5 \Phi$ represents the position of the external force $P$ and moment $T$ along the beam's arc. The non-dimensional variable $\xi$ is:
\begin{equation}
    \xi = \frac{EI}{GK}.
\end{equation}

\noindent $G$ is the shear modulus, and $K$ is the torsional stiffness of a beam with a rectangular section:
\begin{equation}
    K = wt^3\left[\frac{1}{3} - 0.21\frac{t}{w}\left(1-\frac{1}{12}\left(\frac{t}{w}\right)^4\right)\right],
\end{equation}
where $w$ is the width, and $t$ is the thickness of the beam. 

Eq. (\ref{eq:l_delta}) describe the curved beam's deformation as a result of the external force $F$ and moment $T$ at its center, where the non-dimensional functions and constants are defined as:
\begin{equation}
\begin{aligned}
 F_1 &=\frac{1+\xi}{2} \Phi \sin \Phi-\xi(1-\cos \Phi), \\
 F_2 &=\frac{1+\xi}{2}(\Phi \cos \Phi-\sin \Phi) ,\\
 F_3 &=-\xi(\Phi-\sin \Phi)-\frac{1+\xi}{2}(\Phi \cos \Phi-\sin \Phi),\\
F_{a 1} &=\left\{\frac{1+\xi}{2}(\Phi-\Theta) \sin (\Phi-\Theta)-\xi[1-\cos (\Phi-\Theta)]\right\}\langle \Phi-\Theta\rangle^0 ,\\
F_{a 2} &=\frac{1+\xi}{2}[(\Phi-\Theta) \cos (\Phi-\Theta)-\sin (\Phi-\Theta)]\langle \Phi-\Theta\rangle^0.\\
\end{aligned}
\end{equation}

\noindent A complete set of formulations can be found in \cite{young2002roark} Table 9.4.

\newpage

\section*{Supplement Section 6, Experimental Setups for the Haptic Device with the Fluidic Kirigami Metasurface}

To actuate the haptic device and measure its force-displacement response, customized software developed in Python communicates between the computer and data acquisition (DAQ) card (Quanser Q8-USB; Markham, Ontario, Canada). The software operates at a sampling rate of 1kHz to collect data, send commands, and record data for offline analysis. 

Actuation of the haptic device occurs by controlling the pressure commands sent to a high-resolution pressure regulator (Proportion Air QPV; McCordsville, IN, USA). This pressure regulator is equipped with an integrated pressure sensor for closed-loop pressure control. Pressurized air is supplied to the regulator via an air compressor (California Air Tools 1P1060SP; Otay Mesa, CA, USA), while a DC power supply (KEITHLEY Triple-Channel; Beaverton, OR, USA) provides electrical power. The minimum pressure command that the pressure regulator can reliably provide (as shown in Fig.~\ref{fig:haptic}b) is $4.2$kPa. Data from the integrated pressure sensor within the regulator is acquired through the DAQ and read using our custom software.

Experimental displacement measurements from the haptic device are obtained from a fiber optic displacement sensor (PHILTEC RC125; Annapolis, MD, USA). Specifically, the fiber optic sensor measures the displacement of the Kirigami circular contact area as it inflates under applied pressure. The signal from the fiber optic sensor is first amplified and then collected by the custom software through the DAQ. 

To interpret the absolute displacement of the Kirigami circular contact area, calibration data are obtained using a test bench equipped with a laser displacement sensor (Keyence IL-030; Osaka, Japan) (Fig.~\ref{fig:fab}a,b). In this way, the voltage reading from the fiber optic displacement sensor can be compared to the highly accurate displacement magnitude captured by the laser sensor. The laser displacement sensor output gives a repeatability of $1\mu$m after its signal is amplified (Keyence IL-1000; Osaka, Japan), and the data are captured in our custom software through the DAQ. The calibration data were collected when the combined Kirigami and rubber sheets were free to deform (free-stroke data). 

To estimate the force applied to the combined Kirigami and rubber sheet, we could rely on pressure measurements from the pressure sensor. The relationship between the force applied to the combined Kirigami layer and rubber sheet $F$ and the pressure measured by the pressure sensor $P_{measured}$ is modeled as: $F = P_{measured}A_{contact}$, where $A_{contact} \approx 18mm^2$.  

The experimental displacement and force data for the combined Kirigami and rubber sheet are shown in Fig.~\ref{fig:haptic}c.

\begin{figure} [h]
    \centering
    \includegraphics[width=1\linewidth]{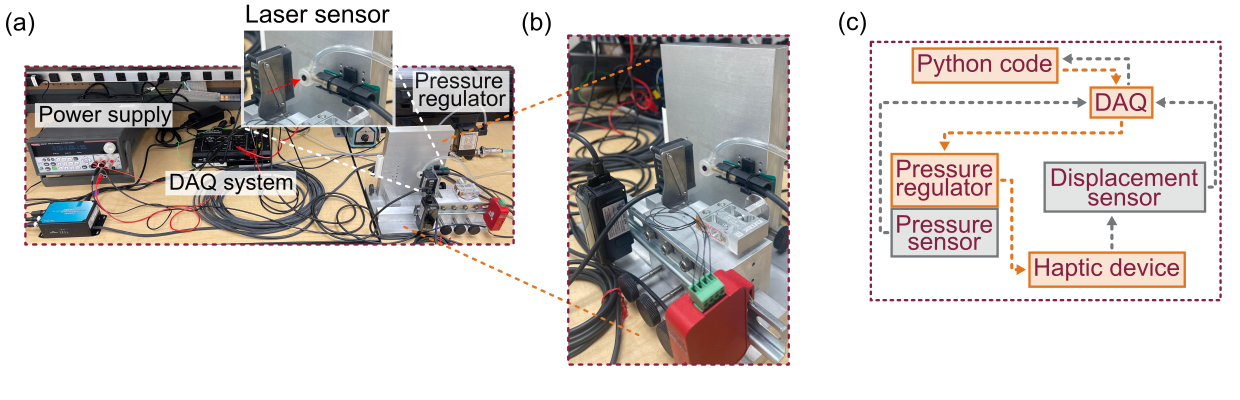}
    \caption{\textbf{Haptic device control system and calibration test bench} (a) The equipment used to operate the haptic device, collect data, and store experimental results. (b) The test bench setup for calibrating the fiber optic sensor using a Keyence displacement sensor. (c) Signal flow diagram summarizing the hardware connections and communications. Gray and orange lines identify components associated with actuating and sensing, respectively.}
    \label{fig:fab}
    \end{figure}

\end{document}